\documentstyle[12pt]{article}
\newcommand{\nc}{\newcommand}
\nc{\renc}{\renewcommand}
\renc{\baselinestretch}{1.1}
\nc{\com}[1]{({\bf {#1}})}

\def\bort#1{ }
\newlength{\overeqskip}
\newlength{\undereqskip}
\nc{\be}[1]{\begin{equation} \mbox{$\label{#1}$}}
\nc{\bea}[1]{\begin{eqnarray} \mbox{$\label{#1}$}}
 \nc{\Section}[2]{\section{\sc #2}\label{#1}\seqnoll}
 \nc{\Subsection}[2]{\subsection{\sc #2}\label{#1}}
\nc{\Bibitem}[1]{\bibitem{#1}}
\nc{\Label}[1]{\label{#1}}
\nc{\eea}{\vspace{\undereqskip}\end{eqnarray}}
\nc{\ee}{\vspace{\undereqskip}\end{equation}}
\nc{\bdm}{\begin{displaymath}}
\nc{\edm}{\end{displaymath}}
\nc{\dpsty}{\displaystyle}
\nc{\bc}{\begin{center}}
\nc{\ec}{\end{center}}
\nc{\ba}{\begin{array}}
\nc{\ea}{\end{array}}
\nc{\bab}{\begin{abstract}}
\nc{\eab}{\end{abstract}}
\nc{\btab}{\begin{tabular}}
\nc{\etab}{\end{tabular}}
\nc{\bit}{\begin{itemize}}
\nc{\eit}{\end{itemize}}
\nc{\ben}{\begin{enumerate}}
\nc{\een}{\end{enumerate}}
\nc{\bfig}{\begin{figure}}
\nc{\efig}{\end{figure}}
\nc{\seqnoll}{\setcounter{equation}{0}}
\renc{\theequation}{\thesection.\arabic{equation}}
\nc{\refc}[1]{\mbox{Ref.~\cite{#1}}}
\nc{\refs}[1]{\mbox{Refs.~\cite{#1}}}
\nc{\eqs}[2]{\mbox{Eqs.~(\ref{#1}) and (\ref{#2})}}
\nc{\eq}[1]{\mbox{Eq.~(\ref{#1})}}
\nc{\figs}[2]{\mbox{Figs.~\ref{#1} and \ref{#2}}}
\nc{\fig}[1]{\mbox{Fig.~\ref{#1}}}

\nc{\figcap}[1]{\begin{quote}\refstepcounter{figure}
        {\bf Figure \thefigure}: {\small #1}\end{quote}}
\nc{\tabcap}[1]{\begin{quote}\refstepcounter{table}
        {\bf Table \thetable}: {\small #1}\end{quote}}

\nc{\tag}[1]{\label{#1} \marginpar{{\footnotesize #1}}}
\nc{\mtag}[1]{\label{#1} \mbox{\marginpar{{\footnotesize #1}}}}
\nc{\etal}{\mbox{\it et al. }}
\nc{\ie}{{\rm i.e. }}
\nc{\eg}{{\it e.g. }}
\nc{\arreq}{&\!\!\!=\!\!\!&}
\nc{\arrmi}{&\!\!\!!-\!\!\!&}
\nc{\arrpl}{&\!\!\!+\!\!\!&}
\nc{\arrap}{&\!\!\!\approx\!\!\!&}
\nc{\non}{\nonumber}
\nc{\nn}{\nonumber\\}
\nc{\align}{\!\!\!\!\!\!\!\!&&}

\nc{\mat}[4]{{\left(\ba{cc} #1 & #2 \\ #3 & #4 \ea\right)}}

\def\simleq{\; \raise0.3ex\hbox{$<$\kern-0.75em
      \raise-1.1ex\hbox{$\sim$}}\; }
\def\simgeq{\; \raise0.3ex\hbox{$>$\kern-0.75em
      \raise-1.1ex\hbox{$\sim$}}\; }

\nc{\DOT}{\hspace{-0.08in}{\bf .}\hspace{0.1in}}
\nc{\Laada}{\hbox {$\sqcap$ \kern -1em $\sqcup$}}
\nc\loota{{\scriptstyle\sqcap\kern-0.55em\hbox{$\scriptstyle\sqcup$}}}
\nc\Loota{{\sqcap\kern-0.65em\hbox{$\sqcup$}}}
\nc\laada{\Loota}
\nc{\qed}{\hskip 3em \hbox{\BOX} \vskip 2ex}

\nc{\real}{{\rm I \! R}}
\nc{\Z}{{\sf Z \!\!\! Z}}
\nc{\complex}{{\rm C\!\!\! {\sf I}\,\,}}

\def\bigid{\leavevmode\hbox{\small1\kern-3.8pt\normalsize1}}
\def\id{\leavevmode\hbox{\small1\kern-3.3pt\normalsize1}}
\nc{\slask}{\hspace{0.1em}\not\hspace{-0.25em}}
\nc{\bis}{{\prime\prime}}
\nc{\pa}{\partial}
\nc{\na}{\nabla}
\def\>{\rangle}
\def\<{\langle}
\def\para{\parallel}
\nc{\goto}{\rightarrow}
\nc{\swap}{\leftrightarrow}
\nc{\EE}[1]{ \mbox{$\times 10^{#1}$} }
\nc{\abs}[1]{\left|#1\right|}
\nc{\at}[2]{\left.#1\right|_{#2}}
\nc{\norm}[1]{\|#1\|}
\nc{\abscut}[2]{\abs{#1}_{\scriptscriptstyle#2}}
\nc{\vek}[1]{\hbox{\boldmath$#1$}}
\nc{\integral}[2]{\int\limits_{#1}^{#2}}
\nc{\inv}[1]{\frac{1}{#1}}
\nc{\dd}[2]{{{\partial #1}\over{\partial #2}}}
\nc{\ddd}[2]{{{{\partial}^2 #1}\over{\partial {#2}^2}}}
\nc{\dddd}[3]{{{{\partial}^2 #1}\over
        {\partial #2 \partial #3}}}
\nc{\dder}[2]{{{d #1}\over{d #2}}}
\nc{\ddder}[2]{{{d^2 #1}\over{d {#2}^2}}}
\nc{\dddder}[3]{{d^2 #1}\over
        {d #2 d #3}}
\nc{\dx}[1]{d\,^{#1}x}
\nc{\dy}[1]{d\,^{#1}y}
\nc{\dz}[1]{d\,^{#1}z}
\nc{\dl}[1]{\frac{d\,^{#1}l}{(2\pi)^{#1}}}
\nc{\dk}[1]{\frac{d\,^{#1}k}{(2\pi)^{#1}}}
\nc{\dq}[1]{\frac{d\,^{#1}q}{(2\pi)^{#1}}}
\nc{\dbar}{d\!\!\!\stackrel{\stackrel{\!-}{}}{}\!\!\!}
\nc{\cc}{\mbox{$c.c.$ }}
\nc{\hc}{\mbox{$h.c.$ }}
\nc{\cf}{cf.\ }
\nc{\erfc}{{\rm erfc}}
\nc{\Tr}{{\rm Tr\,}}
\nc{\tr}{{\rm tr\,}}
\nc{\pol}{{\rm pol}}
\nc{\sign}{{\rm sign}}
\nc{\bfT}{{\bf T }}

\nc{\cA}{{\cal A}}
\nc{\cB}{{\cal B}}
\nc{\cD}{{\cal D}}
\nc{\cE}{{\cal E}}
\nc{\cF}{{\cal F}}
\nc{\cG}{{\cal G}}
\nc{\cH}{{\cal H}}
\nc{\cL}{{\cal L}}
\nc{\cM}{{\cal M}}
\nc{\cO}{{\cal O}}
\nc{\cT}{{\cal T}}
\nc{\1}{\aa}
\nc{\2}{\"{a}}
\nc{\3}{\"{o}}
\nc{\4}{\AA}
\nc{\5}{\"{A}}
\nc{\6}{\"{O}}
\nc{\al}{\alpha}
\nc{\Del}{\Delta}
\nc{\e}{\epsilon}
\nc{\eps}{\epsilon}
\nc{\g}{\gamma}
\nc{\lam}{\lambda}
\nc{\om}{\omega}
\nc{\Om}{\Omega}
\nc{\ve}{\varepsilon}
\nc{\mn}{{\mu\nu}}
\nc{\ka}{\kappa}

\nc{\vp}{\varphi}
\nc{\pub}[4]{\Bibitem{#1}#2, {\sl ``#3''}, #4.}
\nc{\aap}[3]{{\it  Astron.\ Astrophys.\ }{{\bf #1} {(#2)} {#3}}}
\nc{\advp}[3]{{\it  Adv.\ in\ Phys.\ }{{\bf #1} {(#2)} {#3}}}
\nc{\annp}[3]{{\it  Ann.\ Phys.\ (N.Y.)\ }{{\bf #1} {(#2)} {#3}}}
\nc{\annraa}[3]{{\it Ann.\ Rev.\ Astron.\ Astrophys.\ }{{\bf #1} {(#2)} {#3}}}
\nc{\apl}[3]{{\it  Appl. Phys. Lett. }{{\bf #1} {(#2)} {#3}}}
\nc{\apj}[3]{{\it  Ap.\ J.\ }{{\bf #1} {(#2)} {#3}}}
\nc{\apjl}[3]{{\it  Ap.\ J.\ Lett.\ }{{\bf #1} {(#2)} {#3}}}
\nc{\app}[3]{{\it Astropart.\ Phys.\ }{{\bf #1} {(#2)} {#3}}}

\nc{\cmp}[3]{{\it  Comm.\ Math.\ Phys.\ }{{ \bf #1} {(#2)} {#3}}}
\nc{\cqg}[3]{{\it  Class.\ Quant.\ Grav.\ }{{\bf #1} {(#2)} {#3}}}
\nc{\epl}[3]{{\it  Europhys.\ Lett.\ }{{\bf #1} {(#2)} {#3}}}
\nc{\ijmp}[3]{{\it Int.\ J.\ Mod.\ Phys.\ }{{\bf #1} {(#2)} {#3}}}
\nc{\ijtp}[3]{{\it Int.\ J.\ Theor.\ Phys.\ }{{\bf #1} {(#2)} {#3}}}
\nc{\jmp}[3]{{\it  J.\ Math.\ Phys.\ }{{ \bf #1} {(#2)} {#3}}}
\nc{\jpa}[3]{{\it  J.\ Phys.\ A\ }{{\bf #1} {(#2)} {#3}}}
\nc{\jpc}[3]{{\it  J.\ Phys.\ C\ }{{\bf #1} {(#2)} {#3}}}
\nc{\jpg}[3]{{\it J.~Phys.~G:~Nucl.~Part.~Phys.~}{{\bf #1} {(#2)}{#3}}}
\nc{\jap}[3]{{\it J.\ Appl.\ Phys.\ }{{\bf #1} {(#2)} {#3}}}
\nc{\jpsj}[3]{{\it J.\ Phys.\ Soc.\ Japan\ }{{\bf #1} {(#2)} {#3}}}
\nc{\kdmfm}[3]{{\it Kong.\ Dan.\ Mat.\ Fys.\ Med.\ }{{\bf #1} {(#2)} {#3}}}
\nc{\lmp}[3]{{\it Lett.\ Math.\ Phys.\ }{{\bf #1} {(#2)} {#3}}}
\nc{\lncim}[3]{{\it  Lett.\ Nuov.\ Cim.\ }{{\bf #1} {(#2)} {#3}}}
\nc{\mpl}[3]{{\it  Mod.\ Phys.\ Lett.\ }{{\bf #1} {(#2)} {#3}}}
\nc{\naturw}[3]{{\it  Naturwiss.\ }{{\bf #1} {(#2)} {#3}}}
\nc{\ncim}[3]{{\it  Nuov.\ Cim.\ }{{\bf #1} {(#2)} {#3}}}
\nc{\np}[3]{{\it  Nucl.\ Phys.\ }{{\bf #1} {(#2)} {#3}}}
\nc{\pr}[3]{{\it Phys.\ Rev.\ }{{\bf #1} {(#2)} {#3}}}
\nc{\pra}[3]{{\it  Phys.\ Rev.\ }{{\bf A#1} {(#2)} {#3}}}
\nc{\prb}[3]{{\it  Phys.\ Rev.\ }{{\bf B#1} {(#2)} {#3}}}
\nc{\prc}[3]{{\it  Phys.\ Rev.\ }{{\bf C#1} {(#2)} {#3}}}
\nc{\prd}[3]{{\it  Phys.\ Rev.\ }{{\bf D#1} {(#2)} {#3}}}
\nc{\prl}[3]{{\it Phys.\ Rev.\ Lett.\ }{{\bf #1} {(#2)} {#3}}}
\nc{\pl}[3]{{\it  Phys.\ Lett.\ }{{\bf #1} {(#2)} {#3}}}
\nc{\prep}[3]{{\it Phys.\ Rep.\ }{{\bf #1} {(#2)} {#3}}}
\nc{\prsl}[3]{{\it Proc.\ R.\ Soc.\ London\ }{{\bf #1} {(#2)} {#3}}}
\nc{\ptp}[3]{{\it  Prog.\ Theor.\ Phys.\ }{{\bf #1} {(#2)} {#3}}}
\nc{\ptps}[3]{{\it  Prog.\ Theor.\ Phys.\ suppl.\ }{{\bf #1} {(#2)} {#3}}}
\nc{\physa}[3]{{\it  Physica\ A\ }{{\bf #1} {(#2)} {#3}}}
\nc{\physb}[3]{{\it  Physica\ B\ }{{\bf #1} {(#2)} {#3}}}
\nc{\phys}[3]{{\it Physica\ }{{\bf #1} {(#2)} {#3}}}
\nc{\rmp}[3]{{\it  Rev.\ Mod.\ Phys.\ }{{\bf #1} {(#2)} {#3}}}
\nc{\rpp}[3]{{\it Rep.\ Prog.\ Phys.\ }{{\bf #1} {(#2)} {#3}}}
\nc{\sjnp}[3]{{\it Sov.\ J.\ Nucl.\ Phys.\ }{{\bf #1} {(#2)} {#3}}}
\nc{\spjetp}[3]{{\it Sov.\ Phys.\ JETP\ }{{\bf #1} {(#2)} {#3}}}
\nc{\yf}[3]{{\it Yad.\ Fiz.\ }{{\bf #1} {(#2)} {#3}}}
\nc{\zetp}[3]{{\it Zh.\ Eksp.\ Teor.\ Fiz.\ }{{\bf #1} {(#2)} {#3}}}
\nc{\zp}[3]{{\it Z.\ Phys.\ }{{\bf #1} {(#2)} {#3}}}
\nc{\zpc}[3]{{\it Z.\ Phys.\ C\ }{{\bf #1} {(#2)} {#3}}}
\nc{\ibid}[3]{{\sl ibid.\ }{{\bf #1} {#2} {#3}}}
\nc{\Lbm}{{\cal L}^{\beta,\mu}}
\nc{\dLbm}{\Delta{\cal L}^{\beta,\mu}}
\def\vE{{\rm \bf E}}
\def\vB{{\rm \bf B}}

\renewcommand{\baselinestretch}{1.25}
\begin{document}
\thispagestyle{empty}
\begin{flushright}{\begin{tabular}{l}
SUITP-97-16\\
Theoretical Physics Seminar\\
 in Trondheim No 13 1997
 \end{tabular}}
\end{flushright}
\begin{center}
\baselineskip 1.2cm
\vspace{5mm}
{\Huge\bf  Thermally Induced Photon Splitting }
\normalsize
\end{center}
\vspace{10mm}
{\centering
{\large Per Elmfors}\raisebox{1ex}{a}
 and {\large Bo-Sture Skagerstam}\raisebox{1ex}{b}
 \\[5mm]
{\sl \raisebox{1ex}{a}Department of Physics,
   Stockholm University,
 S-113 85 Stockholm, Sweden \\
elmfors@physto.se\\}
{\sl \raisebox{1ex}{b}Department of Physics,
 The Norwegian University of Science and Technology, \\
N-7034 Trondheim,  Norway \\
boskag@phys.ntnu.no\\ } }
%
\vspace{10mm}
\begin{abstract}
\normalsize
\noindent
We calculate  thermal   corrections to the    non-linear QED
effective action for low-energy photon interactions in a background
electromagnetic field. The
high-temperature expansion  shows that at $T \gg m$ the
vacuum contribution is exactly cancelled to all orders in the external
field except for a non-trivial two-point  function contribution. The
high-temperature expansion derived reveals a remarkable cancellation of
infrared sensitive contributions.   As a
result     photon-splitting in the presence of a magnetic field
is suppressed    in  the   presence  of an
electron-positron QED-plasma  at very high temperatures. In a cold and
dense plasma a similar suppression takes place. At the same time
Compton scattering dominates for weak fields and the suppression is
rarely important in physical situations.
\end{abstract}
\newpage
\setcounter{footnote}{0}
\setcounter{page}{1}
%
\begin{center}
  \Section{s:intro}{Introduction}
\end{center}
Recently the non-linear effects induced by virtual electron-positron pairs in
 quantum electrodynamics (QED) have been discussed in great detail in the
literature (\cite{ad96a,ad96b,bai96} and references cited therein)
confirming earlier calculations concerning in particular photon splitting in
external magnetic fields \cite{ad70,ad71}. It is of great interest to notice
that such non-linear QED effects may lead to observable physical effects in
e.g. the  gamma-ray burst  spectra (for reviews see e.g.
\cite{hig+ling91,harding91}) in particular with regard to the so called soft
 gamma-ray repeaters
\cite{baring91,baring+harding95}. Recently,  photon splitting processes
in the magnetosphere of $\gamma$-ray pulsars has also been discussed
 \cite{hbg97},
where it has been argued that
such processes can be comparable to pair-production processes if the magnetic
field $B$ is sufficiently high, i.e. $B \geq B_c$, where $B_c
\approx 4\times 10^{13}$ Gauss is the critical field in QED
 (i.e. $eB_c = m^2$).

In vacuum the presence of virtual
electron-positron pairs leads to the well-known
Euler-Kockel\cite{Euler+Kockel35}--Heisenberg\cite{Heisenberg36}--Weisskopf
\cite{Weisskopf36}--Schwinger\cite{Schwinger51} (EKHWS) effective action,
which was used in the classical paper by
 Adler  on the subject of
computing photon-splitting amplitudes and absorption coefficients in magnetic
fields \cite{ad71}.
Astrophysical models of neutron stars, which have been used in
explanations for the attenuation of $\gamma$-rays, suggest
that photon-splitting
processes in the presence of strong magnetic fields do not necessarily take
place in an environment which may be approximated by a vacuum.
Instead finite temperature $T$ and/or  chemical potential $\mu $ may
have important effects which motivates a study of photon splitting at finite
temperature
(see e.g. \cite{lev+boal85,us+mel95,bu+mil96,es+nar+os96}).
We are considering the limit of weak  fields in order
to see how large the
effects can be and it turns out that for an electrically
neutral $e^+e^-$-plasma in
a weak field the Compton scattering
is dominating whenever the thermally induces splitting amplitude is
appreciable. There is a possibility that the thermal splitting rate
becomes larger than the Compton rate for finite chemical potential and very
low temperature, but our  calculational technique  breaks down in that limit.

It has been noticed in the
literature that if $T \gg m$ all the non-linear terms in the effective
EKHWS action are cancelled by thermal
corrections and the remaining effective action becomes quadratic in the
electro-magnetic field strengths with a non-trivial dependence in $T, \mu $
\cite{bran+frenk95,ElmforsS95,elm+per+ska94,elm+liej+per+ska95}. Below we
reconsider this cancellation explicitly including ${\cal O}(m/T)$
corrections. As a function of the photon energy $\omega$ the results
obtained in the present paper
strictly only apply to the situation where $\omega/m \ll 1$.
In the vacuum sector Adler \cite{ad71} has, however, calculated the exact
$\omega/m$ dependence and verified that this dependence on the
photon splitting processes is rather weak apart from phase-space factors.
The  $\omega/m$-dependence
can therefore be
extrapolated to $\omega/m \simeq 1$, at least for magnetic fields
such that $B/B_c \leq 1$. We expect a similar weak
$\omega/m$-dependence also in the presence of a thermal environment
even though this has not yet been verified explicitly.
Since we are mostly interested in the temperature and magnetic field
dependence we will simply put $\omega = m$.

The paper is organised as follows. In Section
\ref{s:EA} we recall the form of the QED effective action including one-loop
thermal corrections due to the presence of a fermion-antifermion heat-bath.
In Section \ref{s:weak} we perform a weak field expansion of the
effective action
and a high temperature expansion is considered in Section \ref{s:highT}.
In Section
\ref{s:lowT} a similar low temperature expansion is given. A non-covariant
 contribution to the effective action,
fourth order in the external fields, together with sixth order vacuum and
thermal corrections, are considered in Section~\ref{s:split} with regard to
their effect on photon splitting processes.
Final comments are given in Section~\ref{s:grb}.
%
\begin{center}
\Section{s:EA}{Effective Action}
\end{center}
The thermal one-loop non-linear QED effective Lagrangian
${\cal L}^{\beta,\mu}(\vE,\vB)$
for slowly varying electric and magnetic fields
has been calculated \cite{ElmforsS95} with the result
\bea{genform}
        \Lbm(\vE,\vB)&=&  -
        \frac{1}{2\pi^{3/2}}\int_{-\infty}^\infty\frac{dp_0}{2\pi}
        f_F(p_0;A_{0}){\rm Im}\Biggl\{\int_0^\infty \frac{ds}{s}
        e^2ab\cot(esa)\coth(esb)\nonumber \\
        && \times (h(s)-i\epsilon)^{-1/2}
        \exp\left[-i(m^2-i\epsilon)s
        +i\frac{(p_0-eA_0)^2}{h(s)-i\epsilon}+i\frac{\pi}{4}\right]\Biggr\}\ .
\eea
The field-independent contribution of \eq{genform} can be obtained by making
use of dimensional regularisation techniques as described in
\cite{elm+per+ska94}.
The function $h(s)= (eF\coth eFs)_{00}$, where $F_{{\mu\nu}} = \partial_\mu
A_\nu-\partial_\nu A_\mu $, was calculated in
\cite{ElmforsS95} and is
given by
\be{alpha}
        h(s)=
        \left(\inv 2-\frac{{\cal H}}{a^2 + b^2}\right)
        e a\cot\big[esa\big]+\left(\inv 2+\frac{{\cal H}}
        {a^2 + b^2}\right)
        e b\coth\big[esb\big]\ .
\ee
In terms of the field-strength we use the notation
\be{notation}
   {\cal H} = \frac{E^2 + B^2}{2} ~,~~
   a=(\sqrt{{\cal F}^2+{\cal G}^2}+{\cal F})^{1/2} ~,~~
   b=(\sqrt{{\cal F}^2+{\cal G}^2}-{\cal F})^{1/2} ~,
\ee
where the Lorentz-invariant quantities ${\cal F}$ and ${\cal G}$ are given by
\be{invariants}
    {\cal F}= \frac{a^2-b^2}{2} = \frac{B^2-E^2}{2} ,~~
    {\cal G} = ab = {\bf E}\cdot{\bf B}\ ,
\ee
in terms of the magnetic and electric fields ${\bf B}$ and ${\bf E}$
($B=|{\bf B}|$, $E=|{\bf E}|$).  In
\cite{ElmforsS95} it was argued that the fermion equilibrium distribution
function, $f_F(p_0 ,eA_{0})$, in the case of an
 external electromagnetic field, should be chosen in
the following
form
\be{fdef}
       f_F(p_0 ;eA_{0})=\frac{\theta(p_0-eA_0)}{e^{\beta(p_0-\mu)}+1}+
         \frac{\theta(-p_0+eA_0)}{e^{\beta(-p_0+\mu)}+1}\ ,
\ee
where $\beta = 1/T$ is the inverse temperature and $\mu $ the chemical
potential related to
the conserved charge of the system.
In this notation the renormalised low energy
EKHWS vacuum
effective action
for non-linear electromagnetic fields in QED has the well-known form
(see e.g. \cite{itzykson+zuber80}):
\be{Lvac}
    {\cal L}^{\rm vac}(\vE,\vB) =
    -\inv{8\pi^2}\int_0^\infty \frac{ds}{s^3}
    \left[s^2e^2ab\coth(esa)\cot(esb)-1-\frac{s^2e^2}{3}(a^2-b^2)\right]
    e^{-m^2s}\ .
\ee
\begin{center}
\Section{s:weak}{Weak Field Expansion}
\end{center}
The total effective action for a slowly varying background field
is given by
\be{effaction}
        {\cal L}_{\rm eff}(\vE,\vB) = -{\cal F} +
        {\cal L}^{\rm vac}(\vE,\vB) + \Lbm(\vE,\vB)\ .
\ee
The vacuum contribution ${\cal L}^{\rm vac}(\vE,\vB)$ has a straightforward
expansion in terms of
the invariants ${\cal F}$ and ${\cal G}$. We are now interested in a similar
weak-field
expansion of the thermal contribution $\Lbm(\vE,\vB)$. Care
 must then be
taken since, as we will shortly see,  one easily encounters severe infrared
problems.
The basic idea is to collect everything in the integrand in \eq{genform}
which depends on the fields into one function multiplying the zero-field
part and then expand it formally in $\vE$ and $\vB$. For a strictly
degenerate plasma $\Lbm(\vE,\vB)$ is not analytic in the fields but shows
de~Haas--van~Alphen oscillations as discussed in
\cite{elm+per+ska94}. However, at finite temperature and weak enough fields
these oscillations average out over the Fermi surface and a weak field
expansion becomes meaningful. In this situation we can neglect the
$i\epsilon$ in the combination $h(s)-i\eps $ which is only needed for the
non-analytic structure. In particular we need the following
expansion
\be{sexp}
 \frac{esa\cot\big[esa\big]esb\coth\big[esb\big]}{(sh(s)-
i\epsilon)^{1/2}} \exp\left[i\omega ^2 s \frac{1-sh(s)}{sh(s)-i\eps}\right]
= \sum_{k=0}^{\infty} \alpha_{k-2}(\omega;{\cal H};{\cal F};{\cal G})s^k~~,
 \ee
where $\om =p_0-eA_0$ turns out to be  a convenient variable of integration.
The coefficients
$\alpha_{k}(\omega;{\cal H};{\cal F};{\cal G})$ are rather
lengthy and we have most easily evaluated them
using a symbolic computer program. After
performing the
$s$-integrals in \eq{genform}, using the expansion in \eq{sexp}, we
find that
\be{eamitt}
    \Lbm(\vE,\vB)=-\inv{4\pi^{5/2}}\int_{-\infty}^\infty
    d\omega f_F(\om) {\rm Im}\left\{ \sum_{k=-2}^{\infty} \alpha_k(\om)
    \left(-i(\om^2-m^2)+\epsilon\right)^{-\inv2-k}
    \Gamma(k+\inv2)e^{i\frac{\pi}{4}} \right\}\ ,
\ee
where we in $\alpha_k(\om)$ have  suppressed
 the dependence of the variables ${\cal H}$, ${\cal F}$ and
${\cal G}$, and where  we have defined
 $f_F(\om)=f_F(\om + eA_0 ;eA_0 )$. When making use  of the variable $\om $,
the thermal distribution function in \eq{fdef} is naturally expressed in
terms
of the
effective chemical potential $\mu _{\rm eff} = \mu - eA_{0}$.
If we in \eq{eamitt} took the $\eps\goto 0$ limit at this point
we would get serious artificial
infrared problems with the $\om$-integral. We avoid these divergences by
first performing a number of partial integrations using
\be{Aid}
    \left(-i(\om^2-m^2)+\epsilon\right)^{-\inv2-k}=
    \frac{\pi^{1/2}}{i^k\Gamma(k+\inv2)}
    D^k_{\om^2}\left(-i(\om^2-m^2)+\epsilon\right)^{-1/2}~~,
\ee
where $ D_{\om^2}$ is a derivative with respect to $\om^2$.
{}From now on we shall drop the $k=-2$  term since it only
gives  the field independent part (\ie the free energy in the absence of the
external fields) which comes out
in a standard way \cite{elm+per+ska94}. For the rest of
\eq{eamitt},  denoted by $\dLbm$, we rewrite $D^k_{\om^2}$
as derivatives with respect to $\om$ and integrate them all
by parts. This is not a problem since the thermal distribution
function  is zero at the boundaries. We obtain the result (notice that
$\alpha_{k=-1}(\om) =0$):
\be{Lpint}
    \dLbm=-\inv{4\pi^2}\int_{-\infty}^\infty d\om
\frac{\theta(\om^2-m^2)}{\sqrt{\om^2-m^2}}
    \sum_{k=0}^{\infty}i^k\left(\frac{d}{d\om}\inv{2\om}\right)^k
    \left(\alpha_k(\om)f_F(\om)\right)~~,
\ee
where  we have used
${\rm Im} \left[(-i(\om^2-m^2)+\epsilon)^{-1/2}e^{i\frac{\pi}{4}}\right] \goto
\theta(\om^2-m^2)(\om^2-m^2)^{-1/2}$.
The expansion in \eq{Lpint} is only valid if $f_F(\om)$ is smooth
enough. Otherwise, a non-analytic structure like the de~Haas--van~Alphen
oscillation can occur.
We can then write
\be{Lexpansion}
\dLbm=  -\inv{4\pi^2}\int_{-\infty}^\infty d\om
          \frac{\theta(\om^2-m^2)}{\sqrt{\om^2-m^2}}\sum_{n=0}^{\infty}e^{2n}
\Delta ^{(2n)} \Lbm
(\om )\ ,
\ee
where $\Delta ^{(2n)} \Lbm (\om )$ is of the order $2n$
in the
electromagnetic field-strengths $E$ and/or $B$.
The lowest order terms
are given through
\be{exp2}
\Delta ^{(2)} \Lbm (\om ) = -\frac{2f_F(\om)}{3}{\cal F}  +
            \frac{\om\,f_F^{(1)}(\om)E^2}{6}\ ,
\ee
and
\bea{exp4}
 \Delta ^{(4)} \Lbm (\om )  &=&
\frac{f_F(\om)}{60\,\om^4}
           \left(4\cF ^2 +
           7\cG ^2 \right) \nonumber \\
&-&
     \frac{f_F^{(1)}(\om)}{60\,{\om^3}}
           \left( 4\cF ^2 +
           7\cG ^2 \right)
+
     \frac{f_F^{(2)}(\om)}{360\,{\om^2}}
           \left( 16\cF ^2 -8\cF\cH +13\cG ^2
            \right) \nonumber \\
&+&
     \frac{f_F^{(3)}(\om)}{360\,\om}
           \left( 8\cF\cH -8\cF ^2 + \cG ^2 \right) \,
          -\frac{f_F^{(4)}(\om)E^4}{288}\ ,
\eea
where we have used $\cH -\cF = E^2$ and the notation
$f_F^{(n)}(\om)=d^nf_F(\om)/d\om^n$.
%
\begin{center}
\Section{s:highT}{High Temperature Expansion}
\end{center}
In the very high temperature limit, i.e. $T\gg m$, $ |\mu _{\rm eff}|$, we see
 that $f_F(\om)\goto \inv2$ and therefore
only a few terms in the expansion of $\Delta^{(2n)}\Lbm(\om )$ survive. The
quadratic terms go  like $\log(T/m)$ at
high temperature and the higher order corrections go to
constants. Then the $\om$-integral can be performed
 for the higher corrections using
\be{omint}
    \int_{0}^\infty d\om
     \frac{\theta(\om^2-m^2)}{\sqrt{\om^2-m^2}} w^{-k}=
     \frac{\sqrt{\pi}\Gamma(\frac{k}{2})}
     {2m^k \Gamma(\frac{k+1}{2})}~~,
\ee
where $k>0\,$.
In this limit the effective Lagrangian then becomes
\bea{LhTlim}
     \dLbm &=&-\frac{e^2}{12\pi^2}(E^2-B^2)\ln (\frac{T}{m})
     +\frac{e^2E^2}{24\pi^2}\nn
     &&-\frac{e^4}{360\pi^2m^4}[(E^2-B^2)^2+7(\vE\cdot\vB)^2]\nn
     &&-\frac{e^6}{1260\pi^2m^8}[(E^2-B^2)\left(2(E^2-B^2)^2
       +13(\vE\cdot\vB)^2\right)] + \cO (F^8) ~~,
\eea
where $F$ generically stands for an external field.
It is interesting to note that the higher order corrections are exactly equal,
but with opposite sign, to the vacuum corrections obtained from \eq{Lvac}:
\be{Lvac4}
    {\cal L}^{\rm vac}(\vE,\vB)=
        \frac{(a^2-b^2)^2+7 (ab)^2}{360m^4\pi^2}
    +\frac{(b^2-a^2)(2(b^2-a^2)^2+13(ab)^2)}{1260m^8\pi^2} +\cO(F^8) ~~.
\ee
As a matter of fact, by inspection of \eq{genform} and \eq{Lvac}, this
cancellation is true to all orders in the external fields except for quadratic
terms exhibited in \eq{LhTlim} \cite{ElmforsS95}. In passing, we also notice
that at zero temperature and for a large
chemical potential,
$|\mu_{\rm eff}| \gg m$, the thermal non-linear effective action
${\cal L}^{\beta,\mu}(E,B)$
also cancels  the vacuum effective action ${\cal L}^{\rm vac}(E,B)$.

At high temperature the vacuum and thermal effects thus cancel
and we have to go to higher order in $m/T$ to find the leading
contribution. If we first subtract the $f_F(\om)=\inv2$ part
and the scale out a factor of $T^{-2n-2}$ in the remaining
$\cO(F^{2n})$ terms, we find that the integrand is infrared
finite as $m/T\goto 0$. This is far from obvious and comes
from a remarkable cancellation of different infrared sensitive terms. To
get the
first sub-leading term we thus take the limit $m/T\goto 0$ and then we
are left with a dimensionless integral.

To be more explicit let us take $\mu_{\rm eff}=0$ and
write the subtracted $\cO(F^4)$ Lagrangian as
\bea{LO4}
    \Delta\cL^{(4)}=-\inv{2\pi^2}\int_0^\infty\frac{dp}{\om}
    \align\left[\frac{C_0}{\om^4}(f_F(\om)-\inv2)
      +\frac{C_1}{\om^3}D_\om f_F(\om)
      \right.\nn\align\left.
      +\frac{C_2}{\om^2}D_\om^2 f_F(\om)
      +\frac{C_3}{\om}D_\om^3 f_F(\om)
      +C_4 D_\om^4 f_F(\om)\right]~~,
\eea
where now $f_F(\om)=(e^{\beta\om }+1)^{-1}$ and $p=\sqrt{\om ^2 - m^2}$.
The full expression is perfectly finite, due to cancellation mechanism
alluded to
above,
but each term diverges at $\beta\om=0$. To be able to treat the terms
separately
we use a zeta-function regularisation technique by multiplying
the integrand
in \eq{LO4}
by $\om ^\nu$, considering
$\nu$ large enough first, and then take the limit $\nu\goto 0$ at the end of
the calculation.
Let us therefore  define the function
\be{Idef}
   I(z,n)\equiv\int_0^\infty\frac{dx\,x^{-z}}{(e^x+1)^n}~~,
\ee
and notice that
\bea{Irec}
    I(z,n)&=&I(z,n-1)+\frac{z}{n-1}I(z+1,n-1)~~,\nn
    I(z,1)&=&\frac{1-2^z}{2^z}\frac{\pi^{1-z}\zeta(z)}
    {\sin({\dpsty z\pi /2 )}}~~.
\eea
For positive $z$ the function $I(z,n)$ has been defined in terms of an
analytical
continuation using the Riemann $\zeta$-function.
The integrals in \eq{LO4} can then be performed by partial integrations,
where the boundary terms vanish for large enough $\nu$, and thereby
reducing the result to
\bea{L04fin}
    \Delta\cL^{(4)}&=&
        \frac{31\zeta(5)}{3840\pi^6T^4}
        [B^4+6B^2E^2-12E^2+9(\vE\cdot\vB)^2] ~~.
\eea
The $\cO(F^6)$ correction can be treated in a similar way and we
find
\bea{LO6fin}
        \Delta\cL^{(6)}&=&
       -\frac{511\zeta(9)}{147456\pi^{10}T^8}
        [2B^6+42B^4E^2+36B^2E^4-220E^6
        -7B^2(\vE\cdot\vB)^2\nn
        && +279E^2(\vE\cdot\vB)^2]~~.
\eea
\begin{center}
\Section{s:lowT}{Low temperature expansion}
\end{center}
In many applications, such as photon splitting
(see Section \ref{s:split}), it is
more
interesting to have a low temperature expansion. At high temperature
Compton scattering would anyway dominate over pure splitting (at least
for low $eB/m^2$) and in several
astrophysical situations the temperature is relatively low compared to the
electron mass.
On general grounds we expect low temperature effects to be
exponentially small
since they are Boltzmann suppressed, but as we shall find there is a region
where the effects still are appreciable. The formula we need for the
expansion is for $\mu_{\rm eff} = 0$
\be{wqDpf}
   \int_{0}^\infty d\om
     \frac{\theta(\om^2-m^2)}{\sqrt{\om^2-m^2}} \inv{\om^q}
     \frac{d^p}{d\om^p} f_F(\om)\simeq
        \left\{\ba{ll}
        \dpsty\frac{(-1)^p e^{-m/T}}{2m^qT^p}
        \frac{\Gamma(\inv 2)\Gamma(\frac{q}{2})}{\Gamma(\frac{q+1}{2})}
        &{\rm for}\quad q\geq 1~~,\\[4mm]
        \dpsty\frac{(-1)^p}{m^qT^p}\left(\frac{\pi T}{2m}\right)^{1/2}
        e^{-m/T}&{\rm for}\quad q=-1,0~~.\ea\right.~~
\ee
We find that, although all terms go to zero exponentially, the ones
with more derivatives with respect to $\om$
have higher inverse powers of  $T$ which makes them
relatively large for $T/m\simleq 1$, where also the approximations
in \eq{wqDpf} are numerically good. Using these terms the effective
action to $\cO(F^4)$ becomes
\bea{lTEA}
   \Lbm(\vE,\vB)&\simeq&\frac{e^2}{4\pi^2}\left[\frac{2}{3}
        \left(\frac{2\pi T}{m}\right)^{1/2}\cF
        +\inv 6 \left(\frac{2\pi m}{T}\right)^{1/2}E^2\right]e^{-m/T} \nn
        &&-\frac{e^4}{4\pi^2}\Bigl[
        \frac{4\cF^2+7\cG^2}{45m^4}+\frac{\pi}{120}\frac{4\cF^2+7\cG^2}{m^3T}
        +\frac{16\cF^2-8\cF\cH+13\cG^2}{180m^2T^2}\nn
        &&
        -\frac{\pi}{360}\frac{8\cF\cH-8\cF^2+\cG^2}{mT^3}
        +\frac{E^4}{144T^4}\left(\frac{\pi T}{2m}\right)^{1/2}\Bigr]e^{-m/T}~~.
\eea
\begin{center}
\Section{s:split}{Weak field photon splitting}
\end{center}
In the classic paper by Adler \cite{ad71} the QED photon splitting
process in a background
magnetic field is treated in great detail. One important observation in that
paper is that the $\cO(F^4)$-term does not contribute to the  splitting
amplitude in vacuum.
This result relies on the covariant form of the effective action,
\ie that only $\cF$ and $\cG$ occurs in the explicit expression, and that
all the photons are collinear due to kinematics.
Non-linear effects on the dispersion relation
either  make it possible to have a non-zero opening angle between the split
photons or prohibits the splitting altogether, depending on the polarisation
states.
At finite temperature the situation is, however, different since now
the manifest Lorentz invariance is
broken in the effective action by the  dependence on $\cH$. The consequence
is that photon splitting is allowed already from the
$\cO(F^4)$-term even for collinear photons.

Thermal corrections also alter the dispersion relation of the
photons which changes the kinematic conditions for splitting, and then
not all combinations of photon polarisation  are
possible.  These effects were studied in \cite{ad71} on a general basis and
it was found that the only  allowed process is
$\parallel\goto\perp_{1}+\perp_{2}$, provided the electron density is
not too large ($n_e \simleq 10^{19} {\rm cm}^{-3}$).
Thermal corrections to the dispersion
relation have several origins. First, there is the plasma mass which
is independent of the background field. The static version, the Deb\"ye
mass, can found from \eq{genform} by noting that the $\cO(F^{0})$ term
depends on $A_0$ through the thermal distribution function. Expanding
$\cO(A_0^2)$ gives the Deb\"ye mass \cite{ElmforsS95}. Secondly, higher order
terms like the $\cF \cH$-term
change the dispersion relation in a background field.
Details of the dispersion
relation are complicated by the fact that polarisation tensor is
non-analytic in the energy and
momentum. We shall concentrate on situations where the
photon frequency $\omega $ is  smaller than the electron mass so that our
effective action for slowly varying fields is valid. At the same time we
take the temperature and chemical potential to be low so that correction to
the dispersion relation is small and the photon almost light like.
In this way we do not have any large correction to phase-space integrals
which would arise when the thermal mass is important.
To be more concrete we first consider an $e^+e^-$-plasma with
$\mu_{\rm eff}=0$ and $T/m$ sufficiently small. Then we require that
\be{omcond}
        \om_p^2=\frac{2\pi\alpha}{m}\left(\frac{2mT}{\pi}\right)^{3/2}
        e^{-m/T}\ll \om^2,k^2 \simleq m^2~~,
\ee
which in actual numbers means $T/m\ll 4.5$.
It may at first seem inconsistent to assume the effect on the dispersion
relation to be small and still argue that plasma effects on the splitting
amplitude can be important. However, corrections to the dispersion relation
should be compared with the photon energy and momentum, while thermal
splitting amplitudes should be compared with other splitting or scattering
processes. Moreover, higher order terms in the field strength
are more IR-sensitive at low $T$ and therefore relatively larger.

The new ingredient in a background plasma is the $\cF\cH$-term and
the question is whether there is any regime in which the
thermal splitting is important if the temperature is not much higher
than the
electron rest mass.
At low temperature the vacuum
effects  dominate and at high temperature Compton scattering is
overwhelmingly large.

The various processes to be considered depend differently on the photon energy
$\omega $ and the
scattering angle $\theta$ between the direction of the propagation of the
initial
photon and the external magnetic field, but we shall choose for reasons of
simplicity
$\sin\theta=1$ and $\om=m$ in the final numerical
estimates. This choice of $\om$ is
in the upper limit of the validity since our
approximation assumes a low photon frequency (almost
constant fields) and it should be remembered that
the splitting probability goes like $(\om/m)^5$ (see below), and thus
decreases rapidly for smaller $\om$.

In the low photon frequency  limit the amplitude for
the $\parallel\goto\perp_{1}+\perp_{2}$ photon splitting
process is given by
\bea{Amp}
        && M[\para\goto\perp_{1}+\perp_{2}] =
        \omega\omega_{1}\omega_{2}\left\{
        \left( (\hat{\vek{k}}\times\hat{\vek{\epsilon}}^{\parallel})_{i}
        \frac{\pa}{\pa B_{i}}
        + \hat{\vek{\epsilon}}^{\parallel}_{i}\frac{\pa}{\pa B_{i}} \right)
        \right.   \nn
        &&  \times \left.
        \left( (\hat{\vek{k}}\times\hat{\vek{\epsilon}}^{\perp})_{j}
        \frac{\pa}{\pa B_{j}}
        +\left. \hat{\vek{\epsilon}}^{\perp}_{j}\frac{\pa}{\pa E_{j}}\right) ^2
        \right\}{\cal L}_{\rm eff}(\vE,\vB)\right|_{{\dpsty E =0}}~~.
\eea
The photon absorption coefficient is then given by \cite{ad71}
\be{abscoeff}
        \kappa (\para\goto\perp_{1}+\perp_{2})=
        \frac{1}{32\pi\omega^2}\int_{0}^{\omega}d\omega_{1}
        \int_{0}^{\omega}d\omega_{2}
        \delta (\omega -\omega_{1}-\omega_{2})
        |M[\para\goto\perp_{1}+\perp_{2}]|^2 ~~.
\ee
The appropriate vacuum $\cO(F^6)$ splitting probability for the process
$\para\goto\perp_{1}+\perp_{2}$ can be obtained from \eq{Lvac4}:
\be{kolla}
      \frac{\kappa_6^{\rm vac}}{m}=
        \left(\frac{eB}{m^2}\right)^6\left(\frac{\om}{m}\right)^5\sin^6\theta
        \frac{169\alpha^3}{1488375\pi^2}~~.
\ee
The only difference from the
well-known result of \cite{ad71} is that we use units
where $\alpha=e^2/4\pi$ and $F^{\mu\nu}_{\rm Adler}=\sqrt{4\pi}F^{\mu\nu}$.

\begin{figure}[t]
\unitlength=1mm
\begin{picture}(100,105)(0,0)
\includegraphics{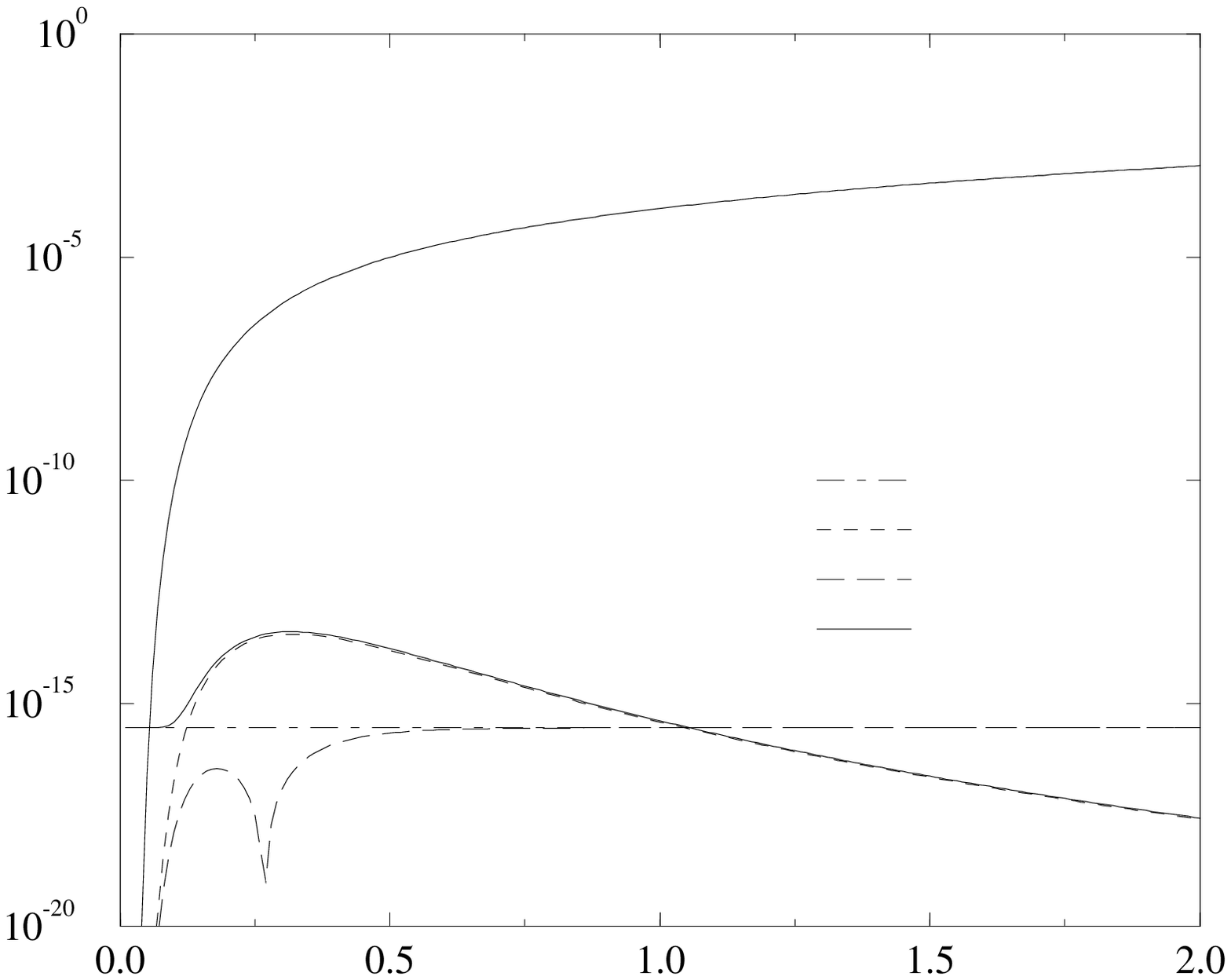}
   \put(78,0){\large $T/m$}
   \put(7,65){\large $\kappa /m$}
     \put(32,93){$eB/m^2=0.2 $}
     \put(32,88){$\omega/m = 1.0$}
     \put(32,83){$\sin \theta = 1.0$}
    \put(110,90){\small$\kappa _C$}
    \put(110,57){\small$\kappa_6^{\rm vac}$}
    \put(110,51.5){\small$\kappa_4^{\beta}$}
    \put(110,46){\small$\kappa_6^{\beta}$}
    \put(110,42){\small$\kappa_{\rm tot}$}
\end{picture}
\figcap{The vacuum and thermal photon splitting  absorption
coefficients  $\kappa_6^{\rm vac}$, $\kappa_4^{\beta}$,
$\kappa_6^{\beta}$, and the Compton absorption $\kappa _C $
(upper solid curve) are shown as a function of
temperature in units of $m$ for $eB=0.2m^2$, $\omega = m$ and
$\sin \theta = 1$.
The lower solid curve corresponds to the coherent sum of amplitudes up to
${\cal O}(F^6)$. By including thermal corrections
of ${\cal O}(F^6)$, the splitting amplitude goes to zero at sufficiently large
$T/m$ as discussed in Section \ref{s:highT}.
\label{f:rate}
}
\end{figure}

Next, we need the splitting probability from the thermal $\cO(F^4)$ term.
The non-covariant part of $\Lbm$ to that order, which is the only one
contributing to photon splitting, is obtained from \eq{exp4} and is given by
\be{O4}
        \Delta\Lbm=-\inv{2\pi^2}\frac{B^4}{360}
        \int_{-\infty}^\infty d\om
        \frac{\theta(\om^2-m^2)}{\sqrt{\om^2-m^2}}
        \frac{d}{d\om}\left(\frac{f^{(2)}(\om)}{\om}\right)~~.
\ee
As in the case of photon-splitting processes in the vacuum,
we derive the splitting rate for the process
$\para\goto\perp_{1}+\perp_{2}$ using of \eq{O4}
\be{k4}
        \frac{\kappa_4^{\beta}}{m}=
        \left(\frac{eB}{m^2}\right)^2\left(\frac{\om}{m}\right)^5\sin^2\theta
        \frac{\alpha^3}{30375\pi^2}
        \left(m^4 \int_0^\infty
        \frac{dk}{\om}
        \frac{d}{d\om}\left(\frac{f^{(2)}(\om)}{\om}\right)\right)^2~~.
\ee
To be consistent we also compute the thermal $\cO(F^6)$ splitting rate
$\kappa^\beta_6$ in a similar way.
Finally it should be noticed that the amplitudes for all
these processes should be added coherently since the final states are the
same and the external field is constant and thus coherent.
The total splitting rate
$\kappa_{\rm tot}$ shows the characteristic decrease for high $T$ when the
thermal contribution interfere destructively with the vacuum part.
The three contributions and the coherent sum are compared in
\fig{f:rate}. (The physical attenuation length in a QED plasma
can be obtained using $m_e^{-1}=3.86\EE{-11}$cm.)

In addition to splitting the photon can scatter directly with the plasma
which turns out to be the dominant process for a large parameter range.
We estimate the scattering rate using the total
Compton cross-section which for unpolarised photons  is
$\sigma_C=8\pi\alpha ^2 f(\omega/m)/3m^2$, where $f(\omega/m)$ is a slowly
varying function such that $f(\om/m)\simeq 1/2$ if $\om \simeq m$
(see e.g. \cite{itzykson+zuber80}). In order
to get the absorption coefficient
$\kappa_C$ we multiply $\sigma_C$
with the appropriate density of electrons and positrons
\be{kC}
        \frac{\kappa_C}{m}=\frac{\sigma_C}{m}
        \frac{2}{\pi^2}
        \int_0^\infty dp\, \frac{p^2}{(e^{\beta\om}+1)}~~.
\ee

It is clear from \fig{f:rate} that the only region where the
thermal splitting can dominate over the vacuum one it is already far below
the Compton scattering rate and will thus not be very important.

\begin{figure}[t]
\unitlength=1mm
\begin{picture}(100,105)(0,0)
\includegraphics{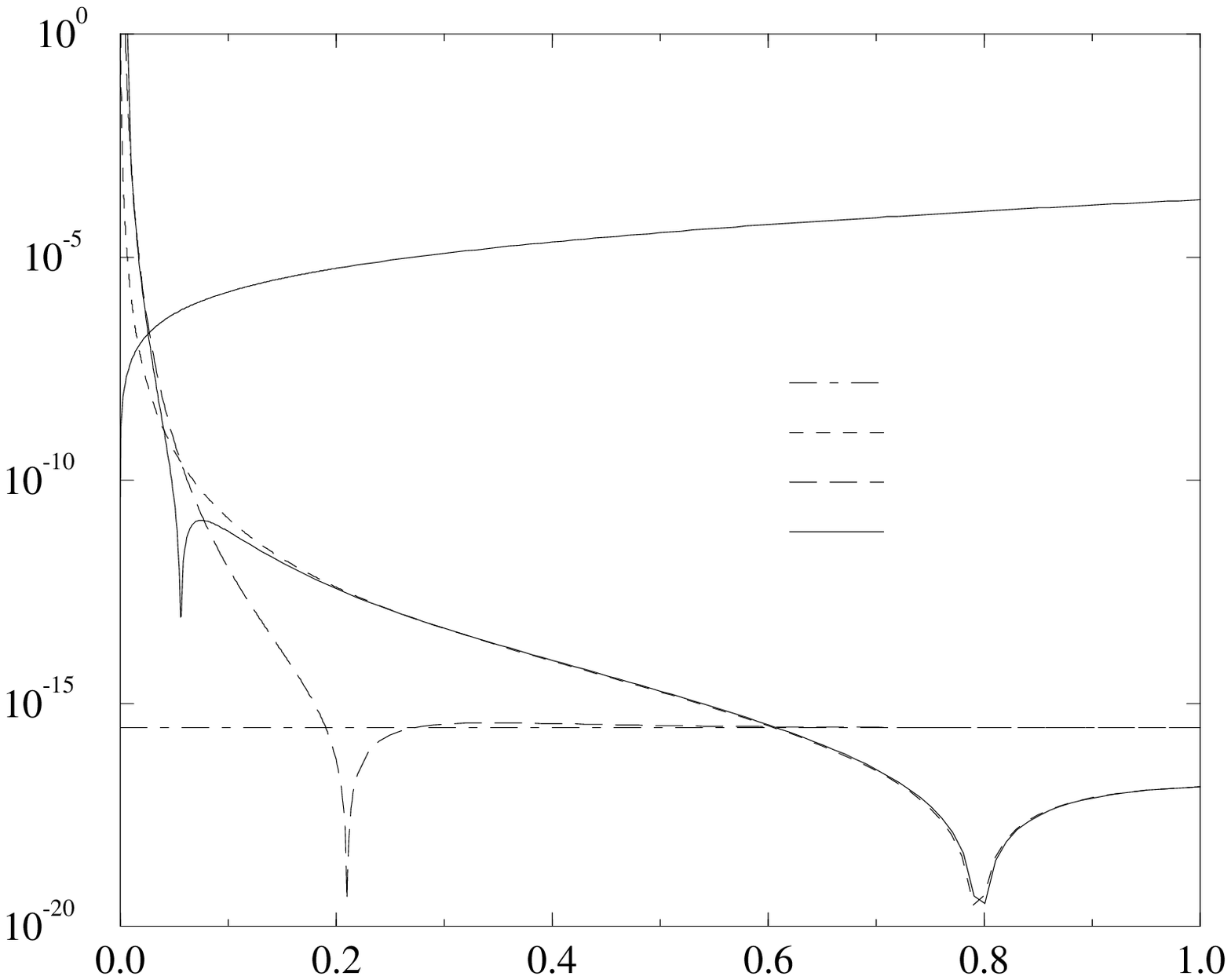}
   \put(78,0){\large $T/m$}
   \put(7,65){\large $\kappa /m$}
     \put(42,70){$eB/m^2=0.2 $}
     \put(42,65){$\omega/m = 1.0$}
     \put(42,60){$\sin \theta = 1.0$}
    \put(101,88){\small$\kappa _C$}
    \put(105,66){\small$\kappa_6^{\rm vac}$}
    \put(105,60.5){\small$\kappa_4^{\beta}$}
    \put(105,55){\small$\kappa_6^{\beta}$}
    \put(105,51){\small$\kappa_{\rm tot}$}
\end{picture}
\figcap{The same symbols and values of parameters as in \fig{f:rate} except
that here the chemical potential is $\mu_{\rm eff}=m$.
\label{f:mu1}
}
\end{figure}

As a second example we shall consider the, for astrophysics,
more realistic case of an electron-proton plasma.
Since the electron is much lighter than the proton
it dominates the correction to the splitting amplitude so we shall only
take into account the electron part. For a non-degenerate plasma at
relatively low temperature ($T\ll m_e$)
the effective chemical potential is close to the
electron mass (larger $\mu_{\rm eff}$ gives a degenerate plasma and smaller
$\mu_{\rm eff}$ corresponds to exponentially small electron density). In
order to see the main features we can therefore put $\mu_{\rm eff}=m_e$.
In \fig{f:mu1} we can see that now the Compton damping rate rapidly
decreases
 for low T while the splitting rates, which are more IR
sensitive, actually increase.
The fact that the thermal splitting
rate becomes large shall in this case
be taken as an indication of that we approach the
limit where the weak field expansion is no longer valid. We know that for
$T=0$ and $\mu_{\rm eff}>m_e$ we have de-Haas--van-Alphen oscillations which
are non-analytic in $B$ and that is reflected in divergent coefficients in
the weak field expansion. Also, $\kappa_6^{\beta}$ diverges more rapidly
than $\kappa_4^{\beta}$ which  shows the breakdown of the perturbative
expansion \cite{elm+per+ska94}. On the other
hand \fig{f:mu1} also tells us that
there might be a region for realistic plasmas where the thermal splitting
actually is the dominant process, but to settle the question it would be
necessary to perform the calculation using the exact Landau levels without
expansion in powers of $B$.
%
\begin{center}
\Section{s:grb}{Final comments}
\end{center}
A detailed analysis of the astrophysical
consequences of thermal photon splitting is outside
the scope of this paper but we have found that in many generic cases this
process is subdominant and not likely to be important.
We have shown in Section
\ref{s:split} that in a QED plasma thermal corrections induce a
non-Lorentz invariant
${\cal O}(F^4)$ term which contributes to the photon splitting
process.
Compared to the ${\cal O}(F^6)$ vacuum contribution this thermal
correction  can actually be quite large already in an $e^+e^-$-plasma
for $T/m\simleq 0.2$, as indicated in \fig{f:rate}, at least if
$eB/m^2 \simeq 0.2$.
For an $e^-p^+$-plasma with chemical potential $\mu_{\rm eff}=m_e$ there is
a possibility of relatively large thermal correction at very low temperature
(see \fig{f:mu1})
but the perturbative calculation in powers of $eB$ breaks down in this
limit. In \fig{f:mu1} the Compton and photon splitting absorption
coefficients become comparable for physical parameters
 corresponding to an electron density of the order of
$10^{28}{\rm cm}^{-3}$.
In a neutron star such an  electron density $n_e$ would occur
only below the surface of the star (the surface electron density can be
the order of $10^{27}{\rm cm}^{-3}$),
 where photon splitting  processes are  not likely to be of any importance
anyhow \cite{hbg97}.
In studies of one-temperature accretion disks
around black holes temperatures like $T\simeq m$ can be
achieved \cite{es+nar+os96}. In such models the magnetic field is,
however, very small ($B\simeq 10^5$G) and photon splitting
is therefore most likely an unimportant physical process. We have not
taken dispersive effects into account. For soft gamma-ray repeaters,
dispersive effects can be of importance  for large electron (and/or
positron) densities in that e.g. the rate for
Compton scattering is modified \cite{bu+mil96}.
%
%
\vspace{5mm}
\begin{center}
{\bf ACKNOWLEDGEMENT}
\end{center}
\vspace{3mm}
The authors thank NorFA for providing partial financial support under
from the NorFA grant 96.15.053-O. P.~E. was also financially supported by
the Swedish Natural Science Research Council under contract 10542-303
and B.-S.~S. was in addition supported  by the Research Council of Norway under
contract
No. 115581/431.
We are grateful to E. \O stgaard
 for providing us with information on gamma ray burst physics.
 \vspace{3mm}
%
%

%
\end{document}